\documentclass[twoside]{article}
\usepackage{amstex}
\evensidemargin=\oddsidemargin \oddsidemargin=1.0cm \evensidemargin=1.0cm
\font\bfit=cmbxti10 
\def\i{\qopname\relax{no}{i}} \def\e{\qopname\relax{no}{e}}
\def\d{\qopname\relax{no}{d}} \def\dsize{\displaystyle}
\def\tsize{\textstyle} \def\SU{\mathop{\rm SU}\nolimits}
\def\SO{\mathop{\rm SO}\nolimits} \def\U{\mathop{\rm U}\nolimits}
\def\ssc#1{\scriptscriptstyle{#1}} \def\noa#1{\noalign{\vskip#1pt}}
\def\alw{\allowdisplaybreaks} 
\def\joinrel{\mathrel{\mkern-3mu}}
\def\Relbar{\mathrel=}
\def\longeq{\Relbar\joinrel=} 
\def\llongeq{\longeq\joinrel=} 
\def\EqRelbar{\mathrel\equiv}
\def\longequiv{\EqRelbar\joinrel\equiv} 
\def\llongequiv{\longequiv\joinrel\equiv} 

\def\yskip{\vskip5pt plus 2pt minus 1pt}
\def\wen#1{$^{[#1]}$} \def\sec#1{\medskip{\noindent #1\par}\yskip}
\def\Td#1#2#3#4#5{{\thispagestyle{empty}
\protect\headheight0pt\protect\headsep0pt\protect\vspace*{-2.2cm}
{\flushleft\parbox{143mm}{\footnotesize Commun.\ Theor.\ Phys.~(Beijing, 
China) {\bf #1}{~(#2)~}{pp~#3}\\[-0.7mm]
\copyright\hspace*{3.5pt} International Academic Publishers\hfill Vol.~{#4},
No.~{#5}}\\[-1.4mm]
\begin{table}[h]\hfill\null\hfill\hrule\vskip.4mm\hrule\end{table} }}}
\def\no#1{\rlap{\protect\rule[-0.25 true cm]{\textwidth}{0.03 true cm}}%
No.~{#1}\hfill}
\def\vo#1{\hfill{\ignorespaces Vol.~{#1}%
\llap{\protect\rule[-0.25 true cm]{\textwidth}{0.03 true cm}}}}
\def\ld{\protect\footnotesize} \def\rd{\protect\footnotesize}
\def\pacs#1{\protect\vglue6pt%
\noindent\begin{minipage}{133mm}{\bf PACS numbers: }\rm #1\end{minipage}}
\def\key#1{\leftskip=1cm\vglue0pt\noindent\hspace*{-4.5pt}\protect
\begin{minipage}{133mm}\begin{minipage}[t]{21mm}{\bf Key words:}\end{minipage}
\hfill\begin{minipage}[t]{111mm}\rm\baselineskip=12pt #1\end{minipage}
\end{minipage}\par\leftskip=0cm}
\def\title#1{\begin{flushleft} 
\Large\bf\protect\baselineskip=17pt #1\end{flushleft}}
\def\author#1{\leftskip=1cm\noindent\begin{minipage}{133mm}
\normalsize\hspace*{-4.5pt}#1\end{minipage}\par\vglue4pt} 
\def\address#1#2{\leftskip=1cm\noindent\begin{minipage}{133mm}\parindent=-4.5pt   
${}^{#1}$\protect\small\baselineskip=10pt #2\end{minipage}\par\vglue2pt} 
 
\def\date#1{\leftskip=1cm\vglue4pt\noindent\begin{minipage}{133mm}
\normalsize\hspace*{-4.5pt}#1\end{minipage}\par\vglue4pt} 
\def\abstract#1#2{\leftskip=1cm\vglue4pt\noindent\hspace*{-4.5pt}\begin{minipage}{133mm}
\small\sl\baselineskip=11pt{\bf Abstract} #1\par
\pacs{\normalsize#2}\end{minipage}\par\vglue2pt\leftskip=0cm}

\begin{document}
\def\fr{\frac}   \def\la{\langle} \def\ra{\rangle} \def\dil{\text{dil.}}  
\def\inver{\text{inver.}} \def\transl{\text{\transl.}} 
\def\Lrightarrow{\mathop{\hbox to 0.8cm{\rightarrowfill}}} \def\C{\text C} 
\abovedisplayskip=4.2pt plus 1pt minus 1pt
\belowdisplayskip=4.2pt plus 1pt minus 1pt
\parskip=0pt plus.2pt minus0.2pt

\setcounter{footnote}{0}
\setcounter{page}{481}
\Td{31}{1999}{481--490}{31}{4, June 15, 1999}  

\title{Pure Spinor Formalism for Conformal Fermion and Conserved Currents}
\author{LIU YuFen$^{1,*}$, MA ZhongQi$^2$ and HOU BoYuan$^{3}$}
\address{1}{Institute of Theoretical Physics, Academia Sinica, P.O.~Box 2735,
Beijing 100080, China}
\address{2}{Institute of High Energy Physics, Academia Sinica, 
P.O.~Box 918(4), Beijing 100039, China}
\address{3}{Department of Physics, The Graduate School at Beijing, 
University of Science and Technology of China, Academia Sinica, 
Beijing 100039, China}
\date{(Received February 8, 1999; Revised March 23, 1999)}

\abstract{Pure spinor formalism and non-integrable exponential
factors are used for constructing the 
conformal-invariant wave equation and Lagrangian density
for massive fermion. It is proved that canonical Dirac Lagrangian for massive
fermion is invariant under induced projective conformal transformations.}
{11.25.Hf}
\key{pure spinor, conformal transformation, triality}

\footnotetext[1]{e-mail: liuyf\@itp.ac.cn}

\baselineskip=13.2pt
\sec{\large\bf I. Introduction}
\vspace*{-0.15cm}
It was proved in 1910 by Bateman and Cunningham\wen{1} that the Maxwell
equations of electromagnetic field are invariant under conformal
transformations in ordinary spacetime. In 1928 Paul Dirac discovered the
relativistic-invariant wave equation of electron. And after several years in
1936 he  studied the possibility of getting a wave equation in conformal
space corresponding to ordinary wave equation for electron.\wen{2} In this
paper we re-examine the above problem by means of Cartan's\wen{3} pure spinor
formalism. One knows that conformal transformations in ordinary spacetime
can be realized as a group of linear $\SO(2,4)$ rotation transformations in a
six-dimensional pseudo-Euclidean space $R^{(2,4)}$. The spinor
representation in $R^{(2,4)}$ exists because the Clifford algebra exists. 
Thus we obtain eight spin-$\frac 12$ 
states rather than the four obtained from the
usual Dirac equation. The physical four-dimensional spacetime manifold is a
``hypersphere'' or a null-cone in an abstract $R^{(2,4)}$. According to
Dirac,\wen{2} only the values of the functions expressed on the four-dimensional
``hypersphere'' represent physical entities. And we conjecture that only 
{\bfit pure spinors} associated with the given isotropic two-plane in 
$R^{(2,4)}$
and the null-geodesic in four-dimensional spacetime have physical
significance. Using the conformal-invariant ``purity-constraint'', we can
write the Lagrangian for fermion in the same form as that for boson, which
is unchanged under conformal rescaling. The mass of fermion was introduced
by the method of introducing the non-integrable exponential 
factor in the fermion. One can realize that the mass term has a close
 relation with ``{\bfit triality}'' in the fermion. We can prove that the usual canonical Lagrangian for
massive fermion is invariant under {\bfit induced} projective conformal
transformations. Finally we give the explicit form of these 
 transformations and the associated Noether's conserved
currents.

\sec{\large\bf II. Conformal Group}
\vspace*{-0.15cm}
Conformal transformations C$(1,3)$ are those (nonlinear) transformations of
the points of (compactified) Minkowski space $x\rightarrow \tilde{x}^\mu 
=\tilde{x}^\mu (x)$ which satisfy the set of equations 
\begin{align}
\frac{\partial \tilde{x}^\mu}{\partial x^\lambda }\;\frac{
\partial \tilde{x}^\nu}{\partial x^\rho }g_{\mu \nu }=\Bigl| \frac{
\partial \tilde{x}}{\partial x}\Bigr| ^{1/2}g_{\lambda \rho } \,,
\end{align}
where $| \frac{\partial \tilde{x}}{\partial x}| $ is the
Jacobian of the transformations. It is the group of transformations which
leaves the equation of line element $\d s^2=0$ invariant. Conformal mappings
include the Poincare rotations-translations $\tilde{x}^\mu =\Lambda _\nu
^\mu x^\nu -h^\mu $ which are metric-preserving, and the simple overall
dilatations $\tilde{x}^\mu\;
\mathop{\llongeq}\limits^{\dil}\;(1+k)x^\mu$. 
The reminder is generated by the involuntary inversions 
\begin{align}
x\,\mathop{\hbox to 1.3cm{\rightarrowfill}}^{\inver}\,\tilde{x}^\mu =x^\mu (x^\nu
x_\nu )^{-1} \,,
\end{align}
which are a four-parameter set since the choice of origin is arbitrary. A
`special conformal transformation' is an inversion, followed by a spacetime
translation followed by an inversion 
\begin{align}
x\,\mathop{\hbox to 1.3cm{\rightarrowfill}}^{\inver}\,\frac{x^\mu }{x^2}
\,\mathop{\hbox to 1.3cm{\rightarrowfill}}^{\text{transl.}}\,
\frac{x^\mu }{x^2}+c^\mu 
\,\mathop{\hbox to 1.3cm{\rightarrowfill}}^{\inver}\,
\frac{x^\mu +c^\mu x^2}{1+2cx+c^2x^2} \;.
\end{align}
These conformal transformations can be realized as a group of linear
transformations in a six-dimensional pseudo-Euclidean space $R^{(2,4)}$. A
clue to this is obtained by noticing that the Lie algebra of C$(1,3)$ is
isomorphic to that of the group $\SO(2,4)$. Introducing pseudo-Euclidean
coordinates $y^{\ssc A}$ (indices $A,B,\cdots=(0,1,2,3,5,6)=(\mu ,5,6)$). The
differential operators $M_{AB}$ representing the generators of the $\SO(2,4)$
are 
\begin{align}
M_{AB}=\i\Bigl(y_{\ssc A}
\frac \partial {\partial y^{\ssc B}}-y_{\ssc B}\frac \partial
{\partial y^{\ssc A}}\Bigr) .
\end{align}

We shall now realize the conformal compactification $M_4^c$ of the Minkowski
space $M_4=R^{(1,3)}$ as a four-dimensional submanifold (with boundary) of 
$R^{(2,4)}$. Consider the con $C$ defined as the set of points in $R^{(2,4)}$
satisfying 
\begin{align}
\eta _{\ssc {AB}}y^{\ssc A}y^{\ssc B}=0 \,,
\end{align}
where $\eta _{\ssc{AB}}=\text{diag}\,(+1,-1,-1,-1,-1,+1)$ is a pseudo-Euclidean metric.
Elements of $M_4^c$ (i.e., equivalence classes of points of $C$) can be
divided into two disjoint subsets.

Let $M=\{y\in C,y^5+y^6\neq 0\}$. We shall identify $M$ with the Minkowski
space $M_4$ and the subset of $M_4^c$ with $y^5+y^6=0$ as the boundary of $
M_4^c$ representation points at infinity. Let us introduce the 
special coordinates on $M$
\begin{align}
x^\mu =\frac{y^\mu }{y^5+y^6} \,.
\end{align}
Then the generators (which act on the hyperquadric $C$) of the conformal
group $C(1,3)$ can be defined as 
\begin{alignat}{3}
& P_\mu &&\equiv  M_{\mu 6}-M_{\mu 5} &&\; =-\i\frac \partial {x^\mu } \,,
\nonumber\\\noa3
& M_{\mu \nu } && \equiv  M_{\mu \nu }
&&\; =\i\Bigl(x_\mu \frac \partial {x^\nu }-x_\nu \frac
\partial {x^\mu }\Bigr) ,\nonumber\\\noa4 
& K_\mu &&\equiv M_{\mu 6}+M_{\mu 5} &&\; =\i\Bigl(x^2\frac \partial {\partial x^\mu
}-2x_\mu x^\nu \frac \partial {\partial x^\nu }\Bigr) ,
\nonumber\\\noa4
& D &&\equiv M_{56}&&\;=\i x^\mu \frac \partial {\partial x^\mu } \,.
\end{alignat}
Strictly speaking it is a restricted conformal group, that is, the subgroup of
mappings connected with the identity map. This does not include the actual
inversion mapping but does include their products with space reflections
(special conformal transformation). The inversion $x\rightarrow \tilde{x}^\mu 
=x^\mu (x^\nu x_\nu )^{-1}$ in $M_4^c$ corresponds to reflection, $
\tilde{y}^6=-y^6$ in $R^{(2,4)}$.

To introduce spinors into our conformal space, we factorize the fundamental
quadratic form $y^{\ssc A}\eta _{\ssc{AB}}y^{\ssc B}$ that defines the space $R^{(2,4)}$. The
spinor representation in $R^{(2,4)}$ exists because the Clifford algebra exists.
The Clifford algebra associated with $\SO(2,4)$ consists of six $8\times 8$
matrices $\Pi _A$ which satisfy 
\begin{align}
\Pi _A\Pi _B+\Pi _B\Pi _A=2\eta _{\ssc{AB}} \,.
\end{align}
The matrices $\hbox{\boldmath$L$}_{AB}=(\i/4)(\Pi _A\Pi _B-\Pi _B\Pi _A)$ generate
 a rotation in the $A$--$B$ plane. They can be considered as generators of 
$\SO(2,4)$ group. This representation is reducible. Specially, in block
notations let 
\begin{align}
\Pi _A=\left( 
\begin{matrix} 
0 & \qquad \Gamma _A \\ 
\Theta _A & \qquad 0 
\end{matrix}
\right) .
\end{align}
Then $4\times 4$ matrices $\Gamma _A$ and $\Theta _B$ satisfy\wen{2} 
\begin{align}
\Gamma _A\Theta _B+\Gamma _B\Theta _A=2\eta _{\ssc{AB}} \,,\qquad 
\Theta _A\Gamma _B+\Theta _B\Gamma _A=2\eta _{\ssc{AB}} \,.
\end{align}
In terms of conventional $4\times 4$ Dirac matrices $\gamma _\mu $ ($\mu
=0,1,2,3$), the special explicit representation is provided by 
\begin{align}
\alw
& \Gamma _A=\left(\matrix 
\Gamma _\mu \\ 
\Gamma _5 \\ 
\Gamma _6 
\endmatrix
\right) =\left( 
\matrix
\dsize \gamma _\mu \Bigl[ 
\sqrt{\frac a{a^{\prime }}}
\Bigl(\frac{1-\gamma _5}2\Bigr)-\sqrt{\frac{a^{\prime }}a}
\Bigl(\frac{1+\gamma _5}2\Bigr)\Bigr] \\\noa4
 \dsize \frac \i{
\sqrt{aa^{\prime }}}\Bigl(\frac{1-\gamma _5}2\Bigr)
+\i\sqrt{aa^{\prime }}\Bigl(\frac{1+\gamma _5}2\Bigr) \\\noa4
\dsize \frac \i{\sqrt{aa^{\prime }}}
\Bigl(\frac{1-\gamma _5}2\Bigr)
-\i\sqrt{aa^{\prime }}\Bigl(\frac{1+\gamma _5}2\Bigr) 
\endmatrix
\right), \nonumber\\\noalign{\vskip12pt}
& \Theta _A=\left( 
\matrix
\Theta _\mu \\ 
\Theta _5 \\ 
\Theta _6 
\endmatrix
\right) =\left( 
\matrix
\dsize \gamma _\mu \Bigl[ 
\sqrt{\frac{a^{\prime }}a}
\Bigl(\frac{1+\gamma _5}2\Bigr)
-\sqrt{\frac a{a^{\prime }}}\Bigl(
\frac{1-\gamma _5}2\Bigr)\Bigr] \\\noa4
\dsize \frac \i{
\sqrt{aa^{\prime }}}\Bigl(\frac{1+\gamma _5}2\Bigr)
+\i\sqrt{aa^{\prime }}\Bigl(\frac{1-\gamma _5}2\Bigr) \\\noa4
\dsize \frac \i{\sqrt{aa^{\prime }}}\Bigl(\frac{1+\gamma _5}2\Bigr)
-\i\sqrt{aa^{\prime }}\Bigl(\frac{1-\gamma _5}2\Bigr) 
\endmatrix
\right).
\end{align}
\vskip 2pt
\noindent 
The matrices $L_{AB}^{-}\equiv (\i/4)(\Gamma _A\Theta _B-\Gamma _B\Theta
_A)$ and $L_{AB}^{+}\equiv (\i/4)(\Theta _A\Gamma _B-\Theta _B\Gamma _A)$
generate different rotations in the $A$--$B$ plane. The group thus has two
``pieces'' $S^{-}=(1+\i\epsilon ^{\ssc{AB}}L_{AB}^{-})\in 
\C_1(1,3)$ and 
$S^{+}=(1+\i\epsilon ^{\ssc {AB}}L_{AB}^{+})\in
\C_2(1,3)$. They are topologically disjunct (disjoint), 
and there is no continuous path from one
piece to the other. On the analogy of Eq.~(7) the above generators of
conformal group which acts on the spinor space can be rewritten in another
form 
\begin{alignat}{2}
\alw 
& P_\mu ^{\pm }  \equiv L_{\mu 6}^{\pm }-L_{\mu 5}^{\pm }  =a^{\pm }\gamma
_\mu \Bigl( \frac{1\pm \gamma _5}2\Bigr),&&\qquad
 L_{\mu \nu }^{\pm }  \equiv L_{\mu \nu }^{\pm }
 =\frac \i4(\gamma _\mu \gamma _\nu -\gamma _\nu \gamma _\mu )\,, 
\nonumber\\\noa3
& K_\mu ^{\pm }  \equiv L_{\mu 6}^{\pm }+L_{\mu 5}^{\pm }  =\frac 1{a^{\pm
}}\gamma _\mu \Bigl( 
\frac{1\mp \gamma _5}2\Bigr) \,,
&&\qquad D^{\pm } \equiv L_{56}^{\pm }  =\pm \frac
\i2\gamma _5 \,,
\end{alignat}
where $a^{-}=a$ , $a^{+}=a^{\prime }$ are constants.

So we can define two types of four-parameter complex semi-spinors $\Phi $ and 
$\Psi $ besides two kinds of gamma matrices $\Gamma _A$ and $\Theta _A$. Under
conformal transformations a pair of spinors transforms as 
\begin{align}
\left( 
\matrix
\tilde{\Phi } \\
 \tilde{\Psi } 
\endmatrix
\right) =\left( 
\matrix
S^{-} & 0 \\ 
0 & S^{+} 
\endmatrix
\right) \left( 
\matrix
\Phi \\ 
\Psi 
\endmatrix
\right) =\left( 
\matrix
(1+\i\epsilon ^{AB}L_{AB}^{-})\Phi \\ 
(1+\i\epsilon ^{AB}L_{AB}^{+})\Psi 
\endmatrix
\right).  
\end{align}
Under inversion $y^{\ssc A}\Pi _A$ must be unchanged, so 
\begin{align}
\left( 
\matrix
\tilde{\Phi } \\ 
\tilde{\Psi } 
\endmatrix
\right) =\left( 
\matrix
0 & \qquad\i\Gamma _6 \\ 
-\i\Theta _6 & \qquad 0 
\endmatrix
\right) \left( 
\matrix
\Phi \\ 
\Psi 
\endmatrix
\right) 
\end{align}
corresponds to reflection $\tilde{y}^6=-y^6$ in $R^{(2,4)}$ and
inversion $x\rightarrow \tilde{x}^\mu =x^\mu (x^\nu x_\nu )^{-1}$ in 
$M_4^c$.

It is easy to prove that \hfill\eject 

\null\vspace*{-1cm}

\begin{align}
& \gamma ^0\Gamma _A^{*T}\gamma ^0=-\Theta _A \,,\nonumber\\\noa3
& \gamma ^0(\Gamma _A\Theta _B-\Gamma _B\Theta _A)^{*T}\gamma ^0=-(\Gamma
_A\Theta _B-\Gamma _B\Theta _A) \,,
\nonumber\\\noa3
& \gamma ^0(\Theta _A\Gamma _B-\Theta _B\Gamma _A)^{*T}\gamma ^0=-(\Theta
_A\Gamma _B-\Theta _B\Gamma _A) \,.
\end{align}
Thus we can define the corresponding conformal conjugate pair of spinors 
$\bar{\Phi }$ and $\bar{\Psi }$ such that $\bar{\Phi }\Phi $
and $\bar{\Psi }\Psi $ are conformal scalars. Their definitions and
transformation properties are 
\begin{alignat}{2}
\alw 
& \bar{\Phi }\,\mathop{\llongequiv}^{\text{def.}}\,\Phi ^{*T}\gamma ^0  \,,
&& \qquad \tilde{\bar{\Phi }}=\bar{\Phi }(S^-)^{-1}=\bar{\Phi }
(1-\i\epsilon^{\ssc{AB}}L_{AB})\,, \nonumber\\\noa3 
& \bar{\Psi }
\,\mathop{\llongequiv}^{\text{def.}}\,\Psi ^{*T}\gamma ^0 \,,
&& \qquad \tilde{\bar{\Psi }}=\bar{\Psi }(S^+)^{-1}=\bar{
\Psi }(1-\i\epsilon ^{\ssc{AB}}L_{AB}^{\prime }) \,.
\end{alignat}

The ``gamma'' matrices $\Gamma _A$ and $\Theta _A$ keep unchanged under the
following
``mixed'' transformations 
\begin{align}
S^{-}(\Gamma _B\Lambda _A^{B})(S^{+})^{-1}=\Gamma _A \,,\qquad 
S^{+}(\Theta _B\Lambda _A^{B})(S^{-})^{-1}=\Theta _A \,,
\end{align}
where $\Lambda_A^B\in \SO\,(2,4)$. 

We can construct two operators $(M_{AB}+L_{AB}^{-})$ and 
$(M_{AB}+L_{AB}^{+}) $, they satisfy the same commutation relations of
conformal Lie algebra. The Casimir operators $C^{-}$ and $C^{+}$ which
commute with operators $(M_{AB}+L_{AB}^{-})$ and $(M_{AB}+L_{AB}^{+})$
are 
\begin{align}
\alw  C^{\pm }\equiv &\;  -\frac{1}4(M_{AB}+L_{AB}^{\pm
})G^{AC}G^{BD}(M_{CD}+L_{CD}^{\pm })
 =  \Bigl[x^\nu \gamma _\nu \gamma ^\mu
-a^{\pm }x^2\i\gamma ^\mu \Bigl( 
\frac{1\pm \gamma _5}2\Bigr)\nonumber\\\noa3
& +2a^{\pm }x^\mu x^\nu \i\gamma _\nu \Bigl(\frac{1\pm \gamma
_5}2\Bigr) 
  +\frac \i{a^{\pm }}\gamma ^\mu \Bigl(\frac{1\mp \gamma _5}2\Bigr)
-2x^\mu \Bigl(
\frac{1\pm \gamma _5}2\Bigr)\Bigr]\frac \partial {\partial x^\mu } \,.
\end{align}

Notice that, these Casimir operators are independent of $\Omega =y^5+y^6$
(conformal scalar of degree zero). They are idempotent operators, i.e., $
(\fr14C^{\pm })(\fr14C^{\pm })=(-\fr14C^{\pm })$. Corresponding
eigenspinors are 
\begin{align}
\Psi _{\text{eig.}}=y^{\ssc A}\Theta _A\phi \,,\qquad 
\Phi _{\text{eig.}}=y^{\ssc A}\Gamma _A\psi \,,
\end{align}
here $y^{\ssc A}$ takes the value only on the spacetime hypersphere. The explicit forms
of these eigenspinors depend upon the choice of a coordinate system. We will
discuss them in the next section.

The metric $g^{\mu \nu }$ induced on $M_4^c$ by the metric $\eta _{\ssc{AB}}$ of 
$R^{(2,4)}$ is given by 
\begin{xalignat}{1}
g^{\mu \nu }= 
\frac{\partial x^\mu }{\partial y^{\ssc A}}\eta^{\ssc{AB}}\frac{\partial x^\nu }{
\partial y^{\ssc B}}=(y^5+y^6)^{-2}\eta ^{\mu \nu }
=\Omega ^{-2}\eta ^{\mu \nu } \,,\qquad 
\sqrt{-g}=(y^5+y^6)^4=\Omega ^4 \,,
\end{xalignat}
where $\Omega =y^5+y^6$. So the
volume element $\sqrt{-g}\,(\d^4x)=\Omega ^4(\d^4x)$ is the conformal scalar of
degree four. For simplicity we consider here the case with a conformal flat
metric only.

A physical theory will be called conformally invariant if it is possible
to attach conformal weights (or degree) to all the quantities appearing in
the theory in such a way that all field equations are preserved under
conformal rescaling. In physics we consider functions of position that can
represent physical entities, such as field quantities or wave functions. 
They are, of course, only the functions of position in the conformal Minkowski
spacetime $M_4^c$.\wen{2} In the next section we will introduce conformal
spinors of degree $n=-1$, i.e., $\Phi =(y^5+y^6)^n\phi (x)$ and $\Psi
=(y^5+y^6)^n\psi (x)$. Where the physical entities $\phi (x)$ and $\psi (x)$
are only functions of $x^\mu $.

\sec{\large\bf III. Pure Spinor}
\vspace*{-0.15cm}

Spinors were first used under such a name by physicists in the field of
quantum mechanics. In most of their general mathematical forms, spinors were
discovered in 1913 by Elie Cartan.\wen{4} For a long
time, the interest of physicists in spinors was restricted to three- and
four-dimensional spaces (Euclidean and Minkowski). Spinors associated with
them have two or four components. Recent work on fundamental interactions
and their unification makes essential use of geometries of more than four
dimensions. For this reason, spinor structure in higher dimensions and, in
particular, Elie Cartan's {\bfit pure} (or simple) spinors, now have 
more chance to relate to physics\wen{5,6} than they had at
the time of the appearance of E. Cartan's lectures.\wen{3} 

We start by schematically reminding some of the main definitions and
properties of pure spinors\wen{3,5,6} associated with a space $V$ of
dimensions $d=6$ with scalar product $g$. Let Cl$(g)$ represent the Clifford
algebra associated with $V$, then a spinor is defined as an eight-dimensional
vector of the complex spinor space $S$ where Cl$(g)$ admits a faithful and
irreducible representation. If $\Pi _1,\Pi _2,\cdots,\Pi _6$ represent an
orthonormal basis of $V$, their elements $\Pi _A$ ($A=1,2,\cdots,6$) may be
thought as generators of Cl$(g)$ with the property
\begin{align}
\Pi _A\Pi _B+\Pi _B\Pi _A=2\eta _{\ssc{AB}} \,.
\end{align}
Let $\boldsymbol\Psi\in S$ be a complex eight-dimensional spinor and
$y\in V$ be a
vector (it may be thought of the form $y=y^{\ssc A}\Pi _A$) satisfying the
equation 
\begin{align}
y{\boldsymbol\Psi =}y^{\ssc A}\Pi _A{\boldsymbol\Psi}=0 \,.
\end{align}
Then for ${\boldsymbol\Psi}\neq 0$, the fundamental quadratic form of $V$ is $
g(y,y)=0$. Therefore we may define $M({\boldsymbol\Psi})$,
\begin{align}
M({\boldsymbol\Psi})=\{y\in V\mid y^{\ssc A}\Pi _A{\boldsymbol\Psi}=0\} \,,
\end{align}
as the subspace of $V$ associated with (the direction of) 
${\boldsymbol\Psi}$. It is easily seen that $M({\boldsymbol\Psi})$ is totally 
null, since from $y,z\in M({\boldsymbol\Psi})$ follows $g(y,z)=0$. We have now:

{\bfit Definition } A spinor ${\boldsymbol\Psi}$ is said to be pure (simple) if 
$M({\boldsymbol\Psi})$ is a maximal and totally null subspace of $V$.

In our case $V$ is {\it real}\/ six-dimensional, the corresponding pure spinor
then defines a 2-plane which must be isotropic (i.e., totally null).
There is Cartan's general theorem:\wen{3}

{\bfit Theorem } Any isotropic $\nu $-plane can be defined in terms of a pure
spinor. The set of pure spinors is left invariant by rotations and reversals.

There are two types of pure spinors, they can be rewritten in the following
different but equivalent forms ($N^{\pm}$ will be defined in Eq.~(27)) 
\begin{alignat}{2}
\alw \Phi &=N^{-}(-a)L_0=\exp \Bigl[-ax^\mu \i\gamma _\mu 
\Bigl(\frac{1-\gamma _5}2\Bigr)\Bigr]L_0 
&& \,\mathop{\llongequiv}^{\text{def.}}\,L \nonumber\\\noa3 
 & =N^{+}\Bigl(\frac 1{ax^2}\Bigr)\Re _0
 =\exp \Bigl[\frac{x^\mu }{ax^2}\i\gamma _\mu 
\Bigl(\frac{1+\gamma _5}2\Bigr)\Bigr]\Re _0 && 
\,\mathop{\llongequiv}^{\text{def.}}\,\Re \nonumber\\\noa3 
 & =N^{-}(-a)L_{01}+N^{+}\Bigl(\frac 1{ax^2}\Bigr)\Re _{02} && 
\,\mathop{\llongequiv}^{\text{def.}}\,L_1+\Re _2
\nonumber\\\noa3 
 & =y^{\ssc A}\Gamma _A\psi &&
\end{alignat}
and 
\begin{alignat}{2}
\alw \Psi & =N^{+}(-a^{\prime })R_0=\exp\Bigl[-a^{\prime }x^\mu \i\gamma _\mu
 \Bigl(\frac{1+\gamma _5}2\Bigr)\Bigr]R_0 && 
\,\mathop{\llongequiv}^{\text{def.}}\,R \nonumber\\\noa3
& =N^{-}\Bigl(\frac 1{a^{\prime }x^2}\Bigr)\ell _0
 =\exp \Bigl[\frac{x^\mu }{a^{\prime }x^2}
\i\gamma _\mu \Bigl(\frac{1-\gamma _5}2\Bigr)\Bigr]\ell _0 && 
\,\mathop{\llongequiv}^{\text{def.}}\,\ell \nonumber\\\noa3
& =N^{+}(-a^{\prime })R_{01}+N^{-}\Bigl(\frac 1{a^{\prime }x^2}
\Bigr)\ell _{02} &&
\,\mathop{\llongequiv}^{\text{def.}}\,R_1+\ell _2  \nonumber\\\noa3
 &=y^{\ssc A}\Theta _A\phi \,.&& {}
\end{alignat} 
We can verify that the eigenspinors of Casimir operators (19)
are pure! We see that the above representations of pure spinors in terms of
components are given for a particular choice of origin and coordinates. To
avoid introducing singularities in the coordinate system, it is convenient
to use the first forms of the above pure spinors. In the next section we will
use these forms. The second forms are useful for observer at infinity, 
here $\Re_0=-ax^\mu \i\gamma _\mu [({1-\gamma _5})/2]L_0$ 
and $\ell _0=-a^{\prime
}x^\mu \i\gamma _\mu [({1+\gamma _5})/2]R_0$. In the third forms $L_{01}+
({x^\mu }/{ax^2})\i\gamma _\mu [({1+\gamma _5})/2]\Re _{02}=L_0$. The $
y^{\ssc A}$ in the last abstract forms takes the value only on the spacetime
hypersphere, and it is not dependent upon the arbitrary choice of origin and
coordinates. We can verify that 
\begin{align}
\bar{\Phi }\Phi =\bar{L}L=\bar{\Re }L=\bar{L}\Re =
\bar{\Re }\Re =0 \,,\qquad \bar{\Psi }\Psi =\bar{R}R=\bar{R}
\ell =\bar{\ell }R=\bar{\ell }\ell =0 \,.
\end{align}
For the study of our pure spinors it is useful to construct a null basis. 
Let
\begin{xalignat}{1}
\hspace*{-1cm}
N^{\pm }(-a^{\pm })  \equiv \Bigl[1-a^{\pm }x^\mu \i\gamma _\mu \Bigl( 
\frac{1\pm \gamma _5}2\Bigr)\Bigr]\Bigl(\frac{1\pm \gamma _5}2\Bigr)  
=\exp \Bigl[-a^{\pm} x^\mu
\i\gamma _\mu \Bigl(\frac{1\pm \gamma _5}2\Bigr)\Bigr]
\Bigl(\frac{1\pm \gamma _5}2\Bigr) 
\end{xalignat}
are idempotent projective operators. They can be considered as
generalization of ordinary $({1-\gamma _5})/2$ and $({1+\gamma
_5})/2$.  We can prove that 
\begin{alignat}{2}
\alw & N^{-}(-a)N^{-}(-a)=N^{-}(-a) \,,&& \qquad  N^{+}
\Bigl(\frac 1{ax^2}\Bigr)N^{+}\Bigl(\frac
1{ax^2}\Bigr)=N^{+}\Bigl(\frac 1{ax^2}\Bigr) \,,\nonumber\\\noa3
& N^{+}\Bigl(\frac 1{ax^2}\Bigr)N^{-}(-a)=N^{-}(-a)\,, && \qquad
 N^{-}(-a)N^{+}\Bigl(\frac
1{ax^2}\Bigr)=N^{+}\Bigl(\frac 1{ax^2}\Bigr) \,.
\end{alignat}
Thus they constitute the algebra of primitive spinorial idempotents and
completely determine the structure of the Clifford algebra and the null
basis of the pure spinors.

In fact the pure spinors 
\begin{align}
\alw
& L\, 
\mathop{\llongeq}^{\text{def.}}N^{-}(-a)L_0=\exp \Bigl[-ax^\nu \i\gamma _\nu 
\Bigl(\frac{1-\gamma _5}2\Bigr)\Bigr]L_0 \,,\nonumber\\\noa3 
& R\,\mathop{\llongeq}^{\text{def.}}N^{+}(-a^{\prime })R_0=\exp 
\Bigl[-a^{\prime }x^\nu
\i\gamma _\nu \Bigl(\frac{1+\gamma _5}2\Bigr)\Bigr]R_0 
\end{align}
are the ``null twistors'' (in four-dimensional spacetime) which were studied
early by Penrose.\wen{7} They can be considered as generalization of
conventional $L_0=[({1-\gamma _5})/2]\psi (x)$ and $R_0=[({1+\gamma _5})/
2]\phi(x)$. The geometry of these spinors is clearest in the terms of
projective twistor space. Considering $\Phi $ to be fixed and solving for
real solutions $x^\mu \in M$ of equation $\Phi =N^{-}(-a)L_0$, it turns out that
a solution exists only if $\bar{\Phi }\Phi =0$. These solutions $x^\mu
(\tau )$ (for fixed $\Phi $) in real Minkowski space $M$ constitute a null
straight line (null geodesics with parameter $\tau $), and every null
straight line in Minkowski space arises in this way. So a point in Minkowski
space is said to be ``incident'' with the null twistor. This is the so-called
standard flat-space twistor correspondence.

We see, our pure spinors (24) and (25) admit a unique decomposition of null
twistors, each of them defines an isotropic two-plane in six-dimensional
space $V$ and a null geodesic (with parameter $\tau $) in physical 
four-dimensional spacetime. In other words, 
for every pure spinor there are two
undetermined parameters in $V$ space, i.e., $\Omega =y^5+y^6$ and $\tau $.

The induced transformations of $L_0\rightarrow \tilde{L}_0=S_0^{-}L_0$
and $R_0\rightarrow \tilde{R}_0=S_0^{+}R_0$ are determined by 
\begin{align}
S^{-}L(x,L_0)  
\equiv L(\tilde{x},\tilde{L}_0)    =\exp \Bigl[-a\tilde{x}^\nu \i\gamma _\nu
\Bigl(\frac{1-\gamma _5}2\Bigr)\Bigr]\{S_0^{-}L_0\} 
\end{align}
and 
\begin{align}
S^{+}R(x,R_0)  \equiv R( 
\tilde{x},\tilde{R}_0)  =\exp \Bigl[-a^{\prime }\tilde{x}^\nu
\i\gamma _\nu \Bigl(\frac{1+\gamma _5}2\Bigr)\Bigr]\{S_0^{+}R_0\} \,.
\end{align}
The most important is although $S^{-}\neq S^{+}$ in Eq.~(13), but after
projection we get $S_0^{-}=S_0^{+}=S_0$. This means that although under
conformal transformations $L$ and $R$ transform in  completely different
ways but after projection, the physical $L_0$ and $R_0$ transform in the
same manner! The infinitesimal generators which correspond to $S_0$ are 
\begin{align}
P_{0\mu }  =0 \,,\quad
L_{0\mu \nu }  =\frac \i4(\gamma _\mu \gamma _\nu -\gamma _\nu \gamma _\mu )\,,
\quad K_{0\mu }  =-\i x^\nu \gamma _\mu \gamma _\nu \,,\quad
D_0  =\frac \i 2 \,.
\end{align}

The induced transformations of $l_0\rightarrow $ $s_0^{-}l_0$ and $\Re
_0\rightarrow s_0^{+}\Re _0$ can be determined in the similar way. We can
prove that infinitesimal generators of $s_0^{-}$ and $s_0^{+}$ are the same,
i.e., 
\begin{align}
P'_{0\mu }  =\i 
\frac{x^\nu }{x^2}\gamma _\mu \gamma _\nu \,,\quad  
L'_{0\mu \nu }  
=\frac\i4(\gamma _\mu \gamma _\nu -\gamma _\nu \gamma _\mu ) \,,\quad 
K'_{0\mu }  =0 \,,\quad 
D'_0  =-\frac \i2 \,.
\end{align}
In quantum theory physical observables must be operators. We can
construct two sets of physical meaningful operators 
$$\alw
\displaylines{\hfill
\left\{\begin{array}{lll}
P_\mu  & =\partial _\mu \,,& {} \\\noa4
M_{\mu \nu }  & =-(x_\mu \partial _\nu -x_\nu \partial _\mu ) 
& \tsize\; -\,\frac14\sigma_{\mu\nu} \,,\\\noa4
 K_\mu  & =2x_\mu x^\lambda \partial _\lambda -x^2\partial _\mu 
& \;+\;x^\nu\gamma_\mu\gamma _\nu\,, \\\noa4
D  & =-x^\mu \partial _\mu  
&\tsize \;-\,\frac 12\,, 
\end{array}\right.
\left\{\begin{array}{lll} 
P'_\mu  & =\partial _\mu 
& \;\dsize-\,\fr{x^\nu}{x^2}\gamma_\mu\gamma_\nu\,, \\\noa4
M'_{\mu \nu }  & =-(x_\mu \partial _\nu -x_\nu \partial _\mu ) 
& \;\tsize -\,\frac14\sigma_{\mu\nu} \,,\\\noa4
 K'_\mu  & =2x_\mu x^\lambda \partial _\lambda -x^2\partial _\mu\,, &{}\\\noa4
D' & =-x^\mu \partial _\mu &\;\tsize +\,\frac 12\,, 
\end{array}\right.
\hfill(34)\cr}
$$
\vskip2pt
\noindent 
where $\sigma_{\mu\nu}=
(\gamma _\mu \gamma _\nu -\gamma _\nu \gamma _\mu )$. The operators
of the first set act on $L_0$ and $R_0$ while operators of the second
set act on $\ell_0$ and $\Re_0$.
Two sets of the above operators satisfy the same commutation relations of
conformal Lie algebra.

Thus, our theory predicts the existence of two kinds of left-handed fermions 
($L_0$ and $\ell _0$) and two kinds of right-handed fermions ($R_0$ and $\Re
_0 $). They have different intrinsic energy-momentum. For observer at
infinity, $K_\mu ^{\prime }$ and $K_\mu $ will play a role of the
energy-momentum. It is easy to see if we replace coordinate $x^\mu $ by 
$z^\mu ={x^\mu }/{x^2}$. Incidentally, one can consider this as a
coordinate change rather than a point transformation.

\sec{\large\bf IV. Conformal-Invariant Dynamical System}
\vspace*{-0.15cm}
A well-known characteristic of weak interactions is that they violate
parity conservation to a maximal degree by virtue of the V--A coupling. That
is, only the left-handed components of leptons are coupled in the
charge-changing sector; the right-handed components play a rather passive
role --- to provide mass. Similarly, hadronic weak interactions can be
accounted for by assuming that quarks have the same kind of weak couplings.
Thus, the elementary entities are states of definite chirality, which have
zero bare mass. An eigenstate of finite mass is a superposition of left- and
right-handed states with equal weights. This is a very important fact
theoretically, because it is the basis for our theoretical understanding of
why quarks and leptons are very light compared with the mass scale of
grand unification or the Planck mass.

The unified gauge theory of electroweak interaction based on a gauge group 
$\SU(2)\times \U(1)$, which {\bfit mixes} different massless chiral states. We
see that in the standard Weinberg--Salam model, the {\bfit left-handed }and 
{\bfit right-handed} \/ fermions are treated on different levels. 
All\/ left-handed
components are supposed to form $\SU(2)$ doublets while the right-handed
components are $\SU(2)$ singlets. This means that from mathematical point of
view the {\bfit left-handed} fermion and the {\bfit right-handed }fermion are
completely different quantities and have different transformation
properties. We will extend this idea to the case with conformal
transformations. In our opinion, there is no reason not to require that 
{\bfit under conformal transformations the left-handed and right-handed fermions
transform in different ways!} We can easily visualize this situation by
the example of two different kinds of screws. When we screw both, the
clockwise ({\it right-handed)} and counterclockwise{\it (left-handed)}
screws, in the same rotation direction, we can observe that if the former is
screwed (translated) in, the latter will be screwed (translated) out.

Generally, the concepts of ordinary {\it left-handed} \/ fermion $L_0\equiv 
[({1-\gamma _5})/2]\phi $ and {\it right-handed} fermion $R_0\equiv 
[({1+\gamma _5})/2]\psi$ are not conformal-invariant. 
For this, the pure spinors
(or the null-twistors)  can be introduced for 
generalization of conventional $L_0$ and $R_0$.

As mentioned in the previous section, in a special coordinate system,
the pure spinors $\Phi $ and $\Psi $ take the form of Eqs~(24) and (25) or 
Eq.~(29). After the redefinition of $L_0$ and $R_0$ 
they can be rewritten in the following null twistor form
\setcounter{equation}{34}
\begin{align}
\alw
& \Phi =  \frac{\Omega ^n}a\exp \Bigl[\i\int (mH_\mu ^{+}-eA_\mu )\d x^\mu -ax^\mu
\i\gamma _\mu \Bigl(\frac{1-\gamma _5}2\Bigr)\Bigr]L_0 \,,\\\noa3
& \Psi =  \frac{\Omega ^n}{
a^{\prime }}\exp \Bigl[\i\int (mH_\mu ^{-}-eA_\mu )\d x^\mu -a^{\prime }x^\mu
\i\gamma _\mu \Bigl(\frac{1+\gamma _5}2\Bigr)\Bigr]R_0 \,.
\end{align}
The additional exponential $\exp [\i\int eA_\mu \d x^\mu ]$ is familiar to us in
the theory of the non-integrable phase factor. The connection between
non-integrability of phase and the electromagnetic field given in this
section is not new, which is essentially just Weyl's principle of gauge
invariance in its modern form.\wen{8} It is also contained in the work of
Ivanenko and Fock,\wen{9} who considered a more general kind 
of non-integrability based on a general 
theory of parallel displacement of half-vectors. The
non-integrable phases for the wavefunctions were also discussed by
Dirac\wen{10} in 1931, where the problem of monopole was studied. 
 C.N. Yang\wen{11} reformulated the concept of a gauge field
 in an integral formalism.
According to Weyl and Yang, the effect of a potential $A_\mu $ is to introduce
a non-integrable phase in the wavefunction $\psi $ of potential-free
particle. Incidentally, following Dirac\wen{2} we assume that electromagnetic
potential $A_\mu $ is a conformal vector of degree $-1$.

The exponential $\exp\{-ax^\mu \i\gamma _\mu 
[({1\pm \gamma _5})/2]\}=\{1-ax^\mu \i\gamma
_\mu [({1\pm \gamma _5})/2]\}$ first appeared in the redefined 
wavefunction of electron in the work of Dirac,\wen{2} 
where the wave equation in conformal space was studied. 
This ``exponential'' was also used in the
definition of null twistor by Penrose\wen{7} and in Ref.~[12].

The exponential $\exp\,[\i m\int H_\mu^{\pm}\d x^\mu]$ 
are introduced here for massive fields, 
$H^\pm_\mu$ must mix different chiral states $
L_0$ and $R_0$. One knows that a charged particle must be massive. Thus in
our model the mass $m$ was introduced in the similar way as the charge $e$. We
give some choice of the ``potentials'' $H_\mu ^{-}$ and $H_\mu ^{+}$ here, 
\begin{align}
\alw & H^{-}\equiv \int H_\mu ^{-}\d x^\mu =\int \Bigl(\frac{\pi _\mu ^{-}}{2
\bar{L}_0R_0}+\frac{\pi _\mu ^{+}}{f^{-}}\Bigr)\d x^\mu \,,
\\\noa3
& H^{+}\equiv \int H_\mu ^{+}\d x^\mu =\int \Bigl(\frac{\pi _\mu ^{+}}{2
\bar{R}_0L_0}+\frac{\pi _\mu ^{-}}{f^{+}}\Bigr)\d x^\mu \,,
\end{align}
here $\pi _\mu ^{-}=\bar{L}_0\gamma _\mu L_0$ and $\pi _\mu ^{+}=
\bar{R}_0\gamma _\mu R_0$. Specially, it is convenient to work with $
H^{-}_\mu=H^{+}_\mu\equiv k_\mu=k_\mu^-+k_\mu^+={\pi_\mu ^{-}}/({2\bar{L}_0
R_0})+{\pi _\mu
^{+}}/({2\bar{R}_0L_0})$. We can verify that $k_\mu^\pm k^{\pm\mu}=0$, 
$k_\mu k^\mu=1$, 
$k_\mu\gamma^\mu R_0=L_0$ and $k_\mu
\gamma^\mu L_0=R_0$. We thus see that there is a curious additional
symmetry referred to as a ``{\bfit principle of
triality}''\wen{3,5,7} with three types of objects: complex vector $k_\mu$,
semi-spinors $R_0$ and $L_0$.  The
most important for us is that if $\psi_0=R_0+L_0$ satisfies massless
Dirac equation, then $\psi=\psi_0\exp\,[-\i m\int k_\mu\d x^\mu]$
will satisfy the {\bfit massive} Dirac equation. 

The simplest 
first order conformally invariant wave equation in six-dimensional space is 
$$
\alw
\displaylines{\hspace*{0.6cm}
\i\Gamma ^A\partial _A\Psi  =W^{-}\i\gamma ^\mu (\partial _\mu -\i eA_\mu
+\i mH_\mu ^{-})R_0 
 =W^{-}[\i\gamma ^\mu (\partial _\mu -ieA_\mu )R_0-mL_0]=0 \,,
\hfill(39)\cr\noa3\hspace*{0.6cm}
\i\Theta ^A\partial _A\Phi  =W^{+}\i\gamma ^\mu (\partial _\mu -\i eA_\mu
+\i mH_\mu ^{+})L_0 
 =W^{+}[\i\gamma ^\mu (\partial _\mu -\i eA_\mu )L_0-mR_0]=0 \,,
\hfill(40)\cr}
$$
here $W^{\pm}=[{-(y^5+y^6)^{-3}}/{\sqrt{aa^{\prime }}}\;]N^{\pm }(-a)\exp
\,(\i mH^{\pm })$ and degree $n=-2$. We realize that the last form of
the above equation is equivalent to the
usual Dirac equation for electron in four-dimensional spacetime, and
its conformal invariance is by no means apparent, but due to its
equivalence to the first form, it of course must also have this
invariant property.

Sometimes it is prefer to use the Lagrangian formalism as a starting point
in constructing various quantum field theories. The point of Lagrangian
formalism is that it makes it easy to satisfy conformal invariance and
especially to obtain Noether's conserved currents.

One knows that the equations of motion follow from a principle of least
action. The action 
\setcounter{equation}{40}
\begin{align}
\boldsymbol{I}=\int \boldsymbol{L}\sqrt{-g}(\d^4x) 
\end{align}
must be unchanged under conformal rescaling. In other words {\bfit the action
must be a conformal scalar of degree zero}, i.e., it must be independent of $
\Omega =(y^5+y^6)$. Since $\sqrt{-g}\,(\d^4x)=\Omega ^4(\d^4x)$ is conformal
scalar of degree $4$, the Lagrangian {\it density } $\boldsymbol{L}(\bar{\psi }
,\psi)$ must be a conformal scalar of degree $-4$. For this, physical
fields must be the conformal fields of degree $n=-1$.

The simplest or most economical conformal-invariant Lagrangian for fermion
in pure spinor's form is (here we take $A_\mu =0$ for simplicity) 
\begin{align}
\alw 
\boldsymbol{L}\sqrt{-g}\d^4x =&\; \Bigl[\Bigl(\frac\partial{\partial 
y^{\ssc B}}\bar{\Phi }\Bigr)G^{BA}\Bigl(\frac \partial {\partial y^{\ssc A}}
\Phi\Bigr)+\Bigl(\frac \partial
{\partial y^{\ssc B}}\bar{\Psi }\Bigr)G^{BA}
\Bigl(\frac \partial {\partial y^{\ssc A}}\Psi \Bigr)\Bigr]
\sqrt{-g}\d^4x \nonumber\\\noa3 
= & \; \Bigl[\Bigl(\frac \partial {\partial x^\mu } 
\bar{\Phi }\Bigr)g^{\mu \nu }\Bigl(\frac \partial {\partial x^\nu }\Phi\Bigr)
+\Bigl(\frac\partial {\partial x^\mu }\bar{\Psi}\Bigr)
g^{\mu \nu }\Bigl(\frac \partial
{\partial x^\nu }\Psi \Bigr)\Bigr]\sqrt{-g}\d^4x \nonumber\\\noa3
 = &\; [\i( \bar{L}_0\gamma _\nu \partial _\mu L_0-\partial _\mu \bar{L}_0
\gamma _\nu L_0)\eta ^{\nu \mu }-2m\bar{L}_0R_0 \nonumber\\\noa3
& +\i(\bar{R}_0
\gamma _\nu \partial_\mu R_0-\partial_\mu \bar{R}_0\gamma _\nu
R_0)\eta ^{\nu \mu }-2m\bar{R}_0L_0]\d^4x \,,
\end{align}
here $\Phi $ and $\Psi $ are pure spinors of degree $n=-1$. The last form of
the above Lagrangian density is nothing but the canonical Dirac's Lagrangian
density of massive fermion. 

In general cases the induced projective transformations of $R_0$
and $L_0$ are very complicated, so for simplicity we will consider another
special  choice of $H^{\pm }$ here. Let $\psi =R_0+L_0$, and 
\begin{align}
H^{\pm }=\int \Bigl(\frac{\bar{\psi }\gamma _\mu \psi }{4\bar{
R}_0L_0}+\frac{\bar{\psi }\gamma _\mu \psi }{4\bar{L}_0R_0}\Bigr)
\d x^\mu=\int \frac{(\bar{\psi }\psi )(\bar{\psi }\gamma _\mu \psi )}
{(\bar{\psi }\gamma_\lambda\psi)\eta^{\lambda\rho}
(\bar{\psi }\gamma_\rho\psi )}\d x^\mu \,.
\end{align}
We want to point out that $H^{+}_\mu=H^{-}_\mu=H_\mu$ are {\it real}  
(real parts of $k_\mu$). 
This redefinition of $
H^{\pm }$ does not change the canonical Dirac's Lagrangian density (42) for
massive fermion. The induced transformations of $R_0$ and $L_0$ are defined
by 
\begin{align}
\alw 
& \tilde{\Phi }=  S^{-}\Phi (x,R_0,L_0,\Omega,\gamma _\mu )  =\Phi ( 
\tilde{x},\tilde{R}_0,\tilde{L}_0,\tilde{\Omega },\gamma
_\mu ) \,,
\nonumber\\\noa3
& \tilde{\Psi }=  S^{+}\Psi (x,R_0,L_0,\Omega ,\gamma _\mu ) 
=\Psi (\tilde{x},\tilde{R}_0,\tilde{L}_0,\tilde{\Omega },\gamma _\mu )\,.
\end{align}
We can prove that $R_0$ and $L_0$ transform in the same way, this
transformation is nonlinear. The induced infinitesimal transformations of $
R_0$ and $\bar{R}_0$ are ($n=-1$ and using Eqs~(12), (13), (16), (30)
and (31)) 
\begin{align}
\alw
 \tilde{R}_0\;\cong\; &\; \Bigl[S_0\Bigl| 
\frac{\partial \tilde{x}}{\partial x}\Bigr| ^{{-n}/4}(\e^{-\i
m\delta H})\Bigr]R_0\quad \cong\quad  \Bigl[1- 
\frac{\varepsilon ^{\mu \nu }}4(\gamma _\mu \gamma _\nu -\gamma _\nu \gamma
_\mu )
\nonumber\\\noa3  
& +c^\nu x^\mu \gamma _\nu \gamma _\mu -\frac k2  
-n(2cx-k)+\i m\int
(2cx-k)H_\mu \d x^\mu \Bigr]R_0 \,,
\\\noa3
\tilde{\bar{R}}_0\;\cong\; &\; \bar{R}_0\Bigl[B_0^{-1}\Bigl| \frac{
\partial \tilde{x}}{\partial x}\Bigr|^{{-n}/4}(\e^{\i m\delta
H})\Bigr] \quad\cong\quad  
\bar{R_0}\Bigl[1+\frac{\varepsilon ^{\mu \nu }}4(\gamma _\mu \gamma
_\nu -\gamma _\nu \gamma _\mu ) \nonumber\\\noa3
 & +x^\nu c^\mu \gamma _\nu \gamma _\mu -\frac
k2 
-n(2cx-k)-\i m\int (2cx-k)H_\mu \d x^\mu \Bigr] .
\end{align}
Notice that $B_0^{-1}S_0=| \fr{\partial \tilde{x}}{\partial x}|
^{{-1}/4}\cong (1+2cx-k)\neq I$. Furthermore we can verify that under
infinitesimal transformations,
\begin{xalignat}{1}
\hspace*{-1cm} \tilde{\Omega }=\Bigl| \frac{\partial \tilde{x}}{\partial x}\Bigr|
^{{-1}/4}\Omega\,, \quad   \tilde{d}^4x=\Bigl| \frac{\partial \tilde{x
}}{\partial x}\Bigr| \d^4x \,,\quad \frac{\partial x^\nu }{\partial \tilde{
x}^\mu }B_0(\gamma _\nu )S_0^{-1}=\gamma _\mu\,, \quad  \tilde{H}_\mu 
=\Bigl| \frac{\partial \tilde{x}}{\partial x}\Bigr| ^{1/4}\frac{
\partial x^\nu }{\partial \tilde{x}^\mu }H_\nu \,,
\end{xalignat}
here $\varepsilon ^{\mu \nu }$, $c^\mu $, $k$ are infinitesimal parameters
of rotation, special conformal transformation and dilation respectively. On
the analogy of the electromagnetic potential, the $H_\mu $ is a
conformal vector of degree $-1$.

Finally we come to the conclusion that the usual canonical action of massive
fermion is invariant under (above) {\bfit induced} 
transformations of conformal group.

There are four conserved Noether's currents associated with the above
induced conformal transformations. The energy-momentum and
angular-momentum currents have canonical usual form 
$$
\alw
\displaylines{\hspace*{1.5cm}
 T_{\nu \cdot }^{\cdot \mu }=\Bigl[\bar{\psi }\i\gamma ^\mu 
\Bigl(\frac
\partial {\partial x^\nu }\psi\Bigr)
-\Bigl(\frac \partial {\partial x^\nu }\bar{
\psi }\Bigr)\i\gamma ^\mu \psi \Bigr] ,
\hfill(48)\cr\noa3\hspace*{1.5cm}
 S_{\nu \lambda }^\mu =  (x_\lambda T_{\nu \cdot }^{\cdot \mu }-x_\nu
T_{\lambda \cdot }^{\cdot \mu }) 
-\frac \i8 
\bar{\psi }[(\gamma _\nu \gamma _\lambda -\gamma _\lambda \gamma _\nu
)\gamma ^\mu +\gamma ^\mu (\gamma _\nu \gamma _\lambda -\gamma _\lambda
\gamma _\nu )]\psi  \,.
\hfill(49)\cr}
$$
The dilation conserved current is 
\setcounter{equation}{49}
\begin{align}
D^\mu =-x^\nu T_{\nu \cdot }^{\cdot \mu }+2(\bar{\psi }\gamma
^\mu \psi )mH \,.
\end{align}
The conserved current associated with special conformal transformation is 
$$
\displaylines{\hfill
C_{\nu \cdot }^{\cdot \mu }=  (2x_\nu x^\lambda -x^2\delta _\nu ^\lambda
)T_{\lambda \cdot }^{\cdot \mu }+x^\lambda 
\bar{\psi }\i(\gamma ^\mu \gamma _\nu \gamma _\lambda -\gamma _\lambda
\gamma _\nu \gamma ^\mu )\psi 
   -4m\bar{\psi }\gamma ^\mu \psi
\int x_\nu H_\lambda dx^\lambda \, .
\hfill(51)\cr}
$$
Notice that,  from equation of motion, for $m\neq 0$ we have $T_{\mu \cdot
}^{\cdot \mu }=2m\bar{\psi }\psi \neq 0$. Using the equation of
motion and conservation equations 
$\partial _\mu T_{\nu \cdot }^{\cdot \mu }=0$ and $
\partial _\mu S_{\nu \lambda }^\mu =0$, we can prove that a sufficient
condition for $\partial _\mu D^\mu =0$ and $\partial _\mu C_{\nu \cdot
}^{\cdot \mu }=0$ is $mH\cdot \partial _\mu (\bar{\psi }\gamma ^\mu
\psi )=0$. The latter is the same with the condition of the $\U(1)$ conserved
current for the $m\neq 0$ case. This conclusion is also true for the $H_\mu
^{-}=H_\mu ^{+}=k_\mu={\pi _\mu ^{-}}/{(2\bar{L}_0R_0)}+{\pi _\mu ^{+}
}/{(2\bar{R}_0L_0)}$ case.

{\bfit Further Speculation } It is interesting to study the change in
$m\oint H_\mu\d x^\mu$ round a closed curve, with the possibility of
there being singularities in $H_\mu$. It leads to quantization of
mass. The detailed discussion of this problem will subject to another
publication.

\vfill
\end{document}